\documentclass[10pt]{article}

\usepackage[left=3cm, right=2cm, top=2cm, bottom=2cm]{geometry}
\usepackage[pdftex]{graphicx}
\graphicspath{Images}
\usepackage{lscape}
\usepackage{makecell}
\usepackage{multirow}
\usepackage{ulem}
\usepackage[utf8]{inputenc}
\usepackage[english]{babel}
\usepackage{amssymb,latexsym,amsmath}
\usepackage{indentfirst} 
\usepackage{multicol}

\usepackage{authblk}
\usepackage[numbers]{natbib}
\usepackage{graphicx}
\usepackage{subcaption}

\title{Polarization singularities in the self-focusing of an elliptically polarized laser beam in an isotropic phase of cholesteric liquid crystal close to the
temperature of phase transition}

\author[1]{G.M. Shishkov}
\author[1,2]{K.S. Grigoriev}
\author[1,2]{V.A. Makarov}
\affil[1]{Faculty of Physics, Lomonosov Moscow State University, Leninskie gory 1 b.2, 119991, Moscow, Russia}
\affil[2]{International Laser Center, Lomonosov Moscow State University, Leninskie gory 1 b.62, 119991, Moscow, Russia}

\date{February 2021}

\begin{document}

\maketitle
\section*{Abstract}
The work is devoted to the self-action of laser beams propagating in the isotropic phase of a cholesteric liquid crystal near the transition temperature to the mesophase in a wide range of parameter values characterizing the nonlocality of the nonlinear optical response of the medium, defocusing in the medium, the power of incident laser radiation and diffraction in the liquid crystal. The origin of the primary singularities of both polarization components of radiation in a wide range of parameters of radiation and the medium is investigated for different values of the parameter characterizing the nonlocality of nonlinear optical response in chiral medium. For each of the values, the visibility diagrams of the primary singularities of both polarization components are plotted depending on the parameters of the medium and radiation, and their asymmetry is found. Dependencies of the radius of noticeable primary C-line and its location in nonlinear medium on the beam and medium parameters.

\section{Introduction}
The aim of this study is to describe the main regularities in the dynamics of polarization characteristics under the self-action of laser beams, as they propagate in the bulk of the isotropic phase of a cholesteric (chiral nematic) liquid crystal (ChLC) at a temperature close to the temperature of nematic-isotropic transition to the mesophase, depending on the values of the parameters characterizing the nonlocality of nonlinear optical response of the medium, defocusing in the medium, incident laser radiation power, dispersion and diffraction in a liquid crystal.
The study of the optical properties of liquid crystals is a promising scientific field. Promising here, for example, is the possibility of obtaining new thin-layer power efficient elements for controlling laser radiation. Recent work in this area has opened up opportunities such as controlling the orbital angular momentum of light using nematic disclinations in liquid crystal films
 \cite{PhysRevLett.111.037802}, creation of electrically tunable
two-dimensional arrays of optical vortices \cite{doi:10.1063/1.4895706, PhysRevLett.121.033901}. 
In the last decades, liquid crystal elements have found their application in the formation of phase and polarization profiles of laser radiation. Similar experiments were carried out in nonlinear optics, for example, due to the Fréedericksz transition in the mesophase of a nematic liquid crystal \cite{ElKetara:12}. For this reason, liquid crystal elements are widely used in experimental singular optics --- the field of optics that studies optical vortices, polarization singularities, and other topological structures of the light field. Previous theoretical studies in the field of interaction between laser pulses and beams containing polarization singularities in the isotropic phase of nematic and cholesteric liquid crystals near the transition to the mesophase have shown the sensitivity of the self-focusing threshold to the polarization of the input radiation; various modes of evolution of the polarization characteristics of the beam during its propagation were obtained. The revealed complex nature of the interaction of circularly polarized radiation components can lead to the appearance of polarization singularities --- special points in the beam cross section, in which the radiation is purely circularly polarized. To date, polarization singularities have been comprehensively investigated in problems of linear optics, but their appearance in nonlinear optical processes remains poorly understood. \cite{grigoriev2011sing, grigoriev2014interaction, grigoriev2015polarization, grigoriev2015formation, grigoriev2019generation}. In this work, we consider the propagation of an initially homogeneously polarized Gaussian beam in the bulk of a cholesteric liquid crystal and describe the main regularities in the dynamics of polarization characteristics during self-focusing.

\section{Numerical model of elliptically polarized beam self-focusing}
Let a monochromatic light beam, propagating along the $Oz$-axis, be incident on the plane boundary of a ChLC lying in plane $z=0$:
\begin{equation}
\mathbf{E}(x,y,z) = \mathbf{A}(x, y, z) \exp[- i\omega t + ikz ]+c.c.,
\end{equation}
where $\omega$ and $k$ are the frequency and a wave number of the beam respectively. For further investigation, it is convenient to represent the slowly varying vector of the complex amplitude of the beam $\mathbf{A}$ as a superposition of two circularly polarized components with right-hand and left-hand polarization. Then the propagation of the beam in the ChLC can be described by a system of nonlinear coupled parabolic equations
\begin{equation}
    \label{SVE}
    \frac{\partial A_\pm}{\partial z} - \frac{i}{2k}\Delta_\perp A_\pm = \frac{2 i\pi k\Delta\chi}{3n^2} (QA_\pm + q_\pm A_\mp),
\end{equation}
written for slowly varying amplitudes $A_\pm = (A_x \pm i A_y)/\sqrt{2}$
of circularly polarized field components. The right-hand sides of the equations \eqref{SVE} include combinations of components of a symmetric traceless tensor $Q_{ij}$ of the ChLC order parameter, namely $Q = Q_{xx}+Q_{yy}$ and $q_{\pm} = Q_{xx}-Q_{yy} \pm 2 i Q_{xy}$,  which, in general, are not equal at different points in the volume of the crystal, and $\Delta \chi$ is the anisotropic term of medium dielectric susceptibility tensor. In the isotropic phase of a ChLC near the phase transition temperature, a strong nonlocality of the nonlinear optical response is observed; therefore, the traditional expansion of the polarization of a substance in powers of the field is inapplicable to describe the optical response of the crystal and is not used in equations \eqref{SVE}. To find combinations of the components of the tensor order parameter $Q$ and $q_\pm$ included in the right-hand sides of these equations, we use the Landau-de Gennes method based on the expansion of the free energy of a liquid crystal in powers of the invariants of the tensor $Q_{ij}$ and its gradient (which is a tensor of the third rank) near the critical temperature. The expression for the free energy density in the isotropic phase of a ChLC, leaving only the leading terms in it, can be written as follows:
\begin{equation}
    \label{freen}
    F = F_0 +\frac{\alpha \Delta T}{2}Q_{ij}Q_{ji}+\frac{L_0}{2}e_{ijk}Q_{im}\partial_kQ_{jm} - \frac{\Delta \chi}{6} Q_{ij} A_i A_j^*.
\end{equation}
In expression \eqref{freen} summing by the repeating indices $i,j,k,m = x,y,z$ is implied, $F_0$ is the free energy density of isotropic ChLC, $e_{ijk}$ is  Levi-Civita tensor, $L_0$ is constant parameter expressing the chiral properties of the crystal, $\Delta T = T-T^*$ is positive difference between the temperature of LC $T$ and the temperature of its transition to the mesophase $T^*$ and $\alpha$ is a temperature-independent constant. Since the considered problem is stationary in time, the search for the minimum of the functional \eqref{freen} was performed using the Euler-Lagrange equations, which for the introduced combinations of $Q$ and $q_\pm$ have the following form:
\begin{equation}
    \label{eulagr}
    \left\{
    \begin{aligned}
        & \alpha \Delta T Q = \frac{\Delta\chi}{18}(|A_+|^2 + |A_-|^2)\\
        & \alpha \Delta T q_\pm \mp i L_0 \frac{\partial q_\pm}{\partial z} = \frac{\Delta\chi}{3}A_{\pm}A_{\mp}^*
    \end{aligned}
    \right.
\end{equation}
Note, that the equation for $Q$ is algebraic, and for $q_\pm$ it is differential. In the present paper, for the second equation of the system \eqref{eulagr}, we chose zero initial conditions ($q_\pm (x, y, 0) = 0$), corresponding to the simplest case of an isotropic cell wall containing a liquid crystal. In turn, the initial conditions of the equations \eqref{SVE} are chosen in the following form
\begin{equation}
    \label{initcond}
    A_{\pm}(x,y,0) = I_0 \sqrt{\frac{1\pm M_0}{2}}\exp\left(-\frac{x^2+y^2}{w^2}\right)
\end{equation}
and describe an elliptically polarized Gaussian beam with the waist size $w$ at $z=0$. In expressions \eqref{initcond} positive parameter $I_0$ defines general beam intensity, and $M_0 \in [-1;1]$ is an ellipticity degree of its polarization ellipses, defined as
\begin{equation}
    M_0 = \frac{|A_+(x,y,0)|^2-|A_-(x,y,0)|^2}{|A_+(x,y,0)|^2+|A_-(x,y,0)|^2}.
\end{equation}
For the further research it is convenient to write equations \eqref{SVE}, \eqref{eulagr} and \eqref{initcond} using the following substitutions
\begin{equation}
\begin{aligned}
    &x_1 = x/w; \quad y_1  = y/w; \quad z_1 = 2z/(kw^2);\\
    &u_\pm = A_\pm/\sqrt{I_0};\quad \tilde{Q} = \frac{18 \alpha \Delta T}{I_0\Delta \chi}Q;\quad \tilde{q}_\pm = \frac{18 \alpha \Delta T}{I_0\Delta \chi}q_\pm,
\end{aligned}
\end{equation}
after which the mathematical model of the problem under study is reduced to the following system of equations
\begin{equation}
    \label{bigsys}
    \left\{
    \begin{aligned}
    &\frac{\partial u_\pm}{\partial z_1} - \frac{i}{4}\tilde\Delta_\perp u_\pm = i P 
    \left[
    \left(|u_+|^2+|u_-|^2\right)u_\pm + \tilde{q}_\pm u_\mp
    \right]\\
    &\tilde{q}_\pm \mp i\mu \frac{\partial \tilde{q}_\pm}{\partial z_1} = 6 u_\pm u_\mp^*\\
    & u_\pm(x_1,y_1,0) = \sqrt{\frac{1\pm M_0}{2}}\exp\left(-x_1^2-y_1^2\right); \quad \tilde{q}_\pm(x_1,y_1,0) = 0
    \end{aligned}
    \right.
\end{equation}
In this system operator $\tilde \Delta_\perp = w^2\Delta_\perp$ and two additional dimensionless parameters are introduced
\begin{equation}
    P = \frac{\pi I_0 (kw\Delta\chi)^2}{54n^2\alpha\Delta t};\qquad \mu = \frac{2L_4}{kw^2\alpha\Delta T},
\end{equation}
the first of which is actually the dimensionless power of the Gaussian beam and expresses the general influence of nonlinear effects during its propagation, and the second is responsible for the nonlocal properties of the ChLC response. Note that the system of equations \eqref{bigsys} is symmetric with respect to the permutation of $u_+$ and $u_-$ and a simultaneous change in the signs of the parameters $M_0$ and $\mu$. This means that a simultaneous change of the initial polarization of radiation and the chirality of the ChLC does not change the self-action of propagating beam. Therefore, without loss of generality, it is assumed below that the parameter $\mu$ is positive. In light of its extreme complexity, the system \eqref{bigsys} can be solved only by numerical methods, and in this work, we used an iterative finite difference method with splitting by physical factors. At each step of the algorithm, the propagation equations and the evolution equation for $\tilde{q}_\pm$ were solved at first without taking into account the operator $\tilde{\Delta}_\perp$ using an explicit-implicit difference scheme and the method of iterative approximations. Then, a linear diffraction problem was solved for the obtained values of $u_\pm$ with modified phases.

\section{Discussion of the results}
\subsection{Self-focusing regime}
\begin{figure}
     \centering
     \includegraphics[width=\textwidth]{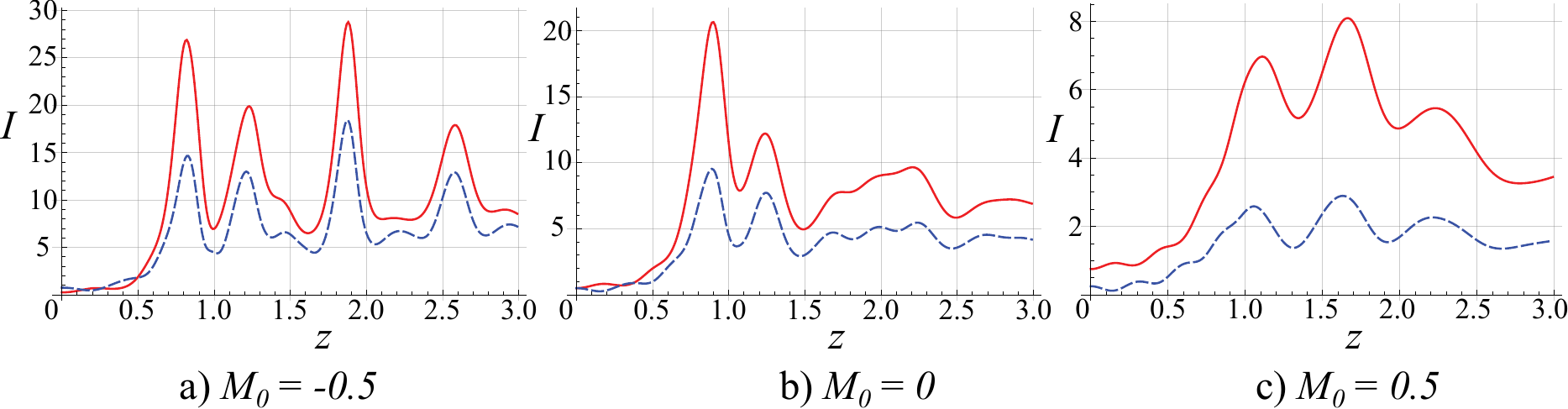}
        \caption{Dependencies of the intensity of right-hand polarized (solid red line) and left-hand polarized (dashed blue line) components in the center of the beam  on the propagation coordinate $z$ for three different values of ellipticity degree of the incident Gaussian beam: a) $M_0=-0.5$, b)$M_0=0$, c)$M_0=0.5$. Propagation coordinate $z$ here and further is normalized by the diffraction length $L_d=kw^2/2$. Other parameters are $P=1$, $\mu = 0.07$.}
        \label{M0}
\end{figure}

Propagating in the bulk of a ChLC, an initially homogeneously elliptically polarized Gaussian beam experiences a nonlinear effect of the medium, which, for the corresponding values of the parameters, can lead to self-focusing of radiation. In the course of modeling, modes were discovered in which the beam propagation has a multifocal character. Such behavior can be explained by the presence of the spatial nonlocality of the nonlinear optical response. The influence of this kind of nonlocality is defocusing. During propagation, the competition between two processes becomes noticeable --- self-focusing in a medium with Kerr-like nonlinearity the defocusing, caused by the presence of spatial dispersion. The competition of these processes can lead to the appearance of a multifocal structure of both polarization components during propagation in a light beam, as shown in fig. \ref{M0}(a)--\ref{M0}(c). A similar behavior was previously observed by the authors in nematic liquid crystal \cite{grigorev2017}. Nevertheless, the propagation of light in ChLC has an essential feature. Figures \ref{M0}(a) -- \ref{M0}(c) show the dependencies of the intensity of both beam components on its axis --- right-hand (solid red line) and left-hand (dashed blue line) for three different values of initial ellipticity degree of the incident Gaussian beam. These values correspond to three qualitatively different cases. Figure \ref{M0}(a) represents the incident beam having an initial ellpiticity degree $M_0$ equal to $-0.5$, which means that the incident radiation has left-handed polarization. It can be seen from the figure that the left-hand polarized component in this case prevails at the boundary of the medium. However, after passing half of the diffraction length, right-handed polarization starts dominating, and this dominance remains stable during its further propagation. Figure \ref{M0}(b) corresponds to the linear polarization of the incident radiation. However, even in this case, after a short period of ``competition'', the right-hand polarized component begins to prevail. In case, shown on fig. \ref{M0}(c) the right-hand polarized component prevails both at the boundary of the medium and throughout the entire length of the beam propagation. Moreover, in all three cases, the ratio between the amplitudes of the components comes to an almost constant value as propagation distance grows. Thus, it turns out that during self-focusing in ChLC, the medium ``prefers'' a certain direction of rotation of the electric field vector, regardless of the polarization of the incident radiation. In addition, this “preference” by the medium of right-hand polarization depends on the sign of the parameter $\mu$, and changes to the opposite as its sign changes.

\subsection{Polarization singularities}

Taking into consideration the polarization of radiation instead of just the intensity, made it possible to detect singular points in individual cross-sections of the beam at which the initially elliptical polarization of the radiation degenerates into circular. The sets of these points in the cross-sections of the beam form circumferences on which the intensity of one of the circularly polarized components abruptly takes zero value. These circumferences are the examples of polarization singularities and in literature are called $C$-lines. From the regularities obtained using the numerical solution, it can be seen that the lines of the polarization singularity of each of the components, as a rule, appear in those beam cross sections corresponding to the $z$ coordinates, in which the peak intensity of considered circularly polarized component attains local maxima or minima. This feature can be applied in real experiment to detect polarization singularities to narrow the search area along the propagation coordinate. There are no similar studies of nonlinear self-action in the isotropic phase of a chiral NLC at a temperature close to the temperature of nematic-isotropic transition.

\begin{figure}
    \centering
    \includegraphics{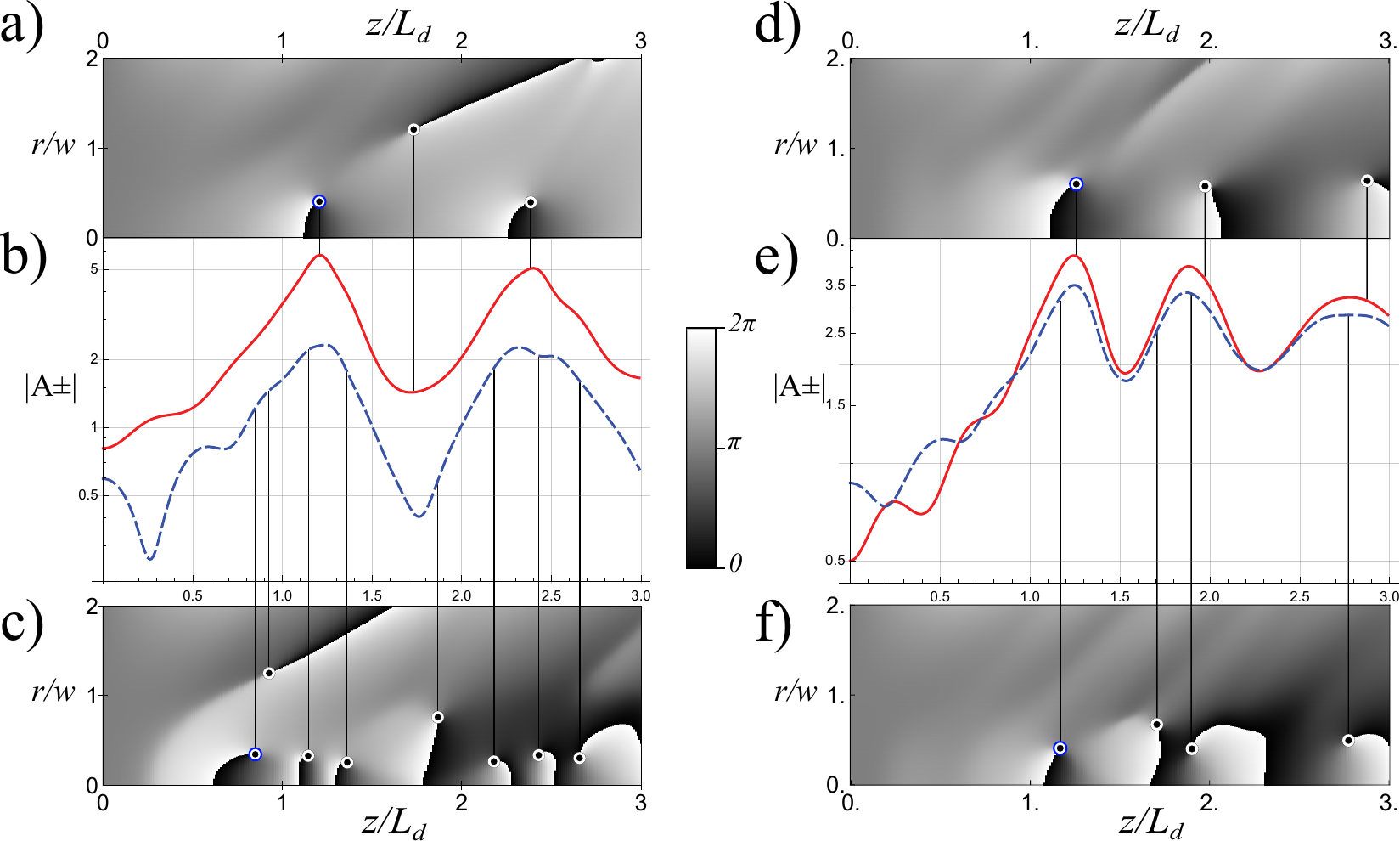}
    \caption{Dependencies of beam properties for the following parameters $P=0.7$, $\mu = 0.07$, $M_0=-0.5$ (a--c) and $P=1.9$, $\mu = 0.3$, $M_0=0.3$ (d--f). Subfigures a, d show the evolution of phase radial distribution of right-hand circularly polarized component and (c, f) show the same of left-hand polarized one. Subfigures b, e (logarithmic scale of the amplitude axis) show dependencies of the right-hand (solid red line) and left-hand (dashed blue line) polarized component amplitudes in the center of the beam. Thin vertical lines designate cross-sections, containing $C$-lines. Their intersections with plane containing $z$-axis are marked by circles in (a, c, d, f). Primary $C$-lines are marked by blue circles. The propagation coordinate $z$ is normalized on beam diffraction length $L_d=kw^2/2$.}
    \label{fig:my_label}
\end{figure}

\subsection{Primary singularities}

\begin{figure}[h!]
\centering
\includegraphics[width=\textwidth]{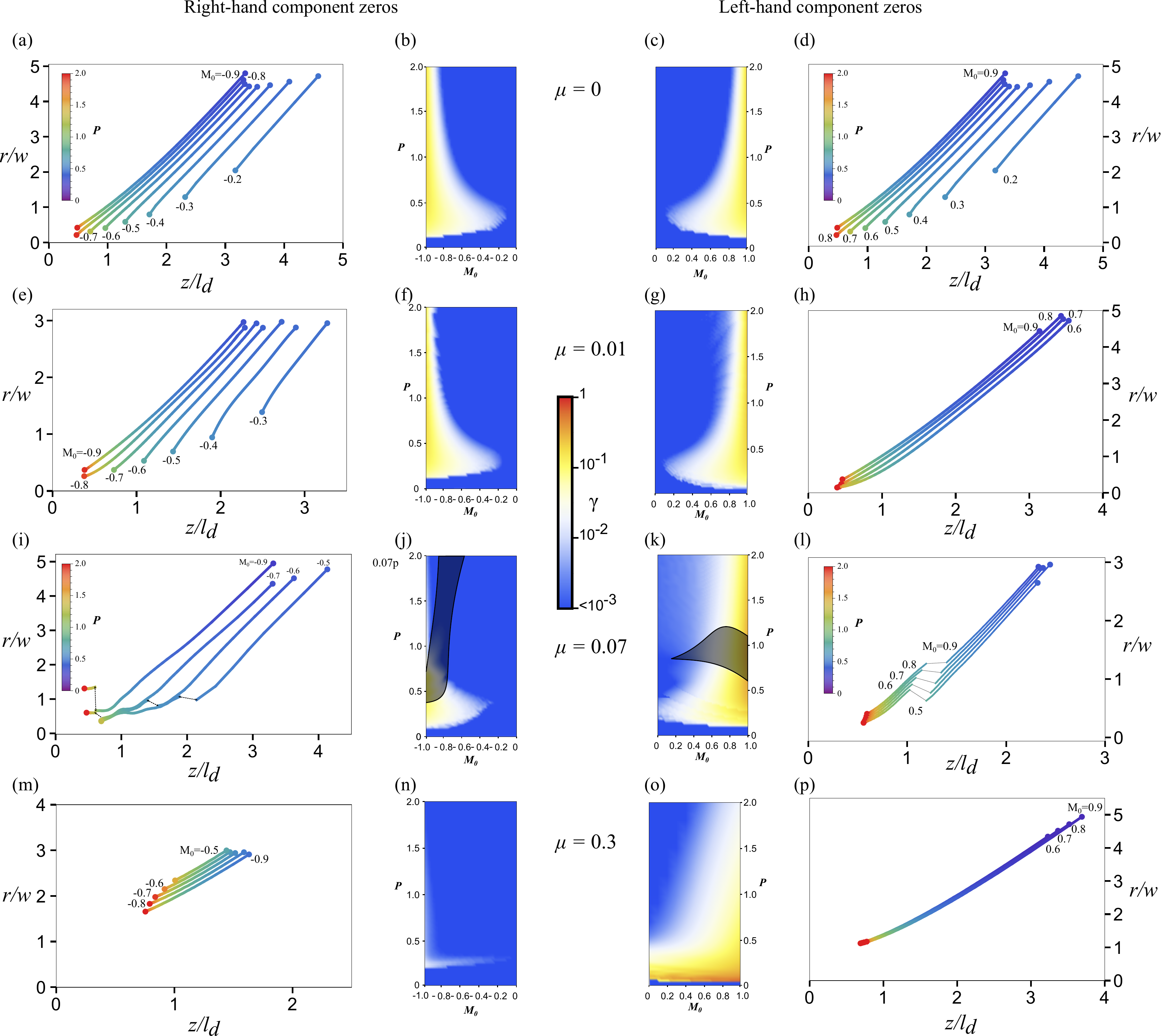}
\caption{Diagrams (b, c, f, g, j, k) of the ratio $\gamma$ of the intensity in the primary singular line and in the center of the beam in the same transverse cross section for different
values of the parameter $P$ and initial ellipticity degree $M_0$ for the primary singularities of the right-hand (b, f, j, n) and left-hand (c, g, k, o) polarization components. Dependencies (a, d, e, h, i, l, m, p) of the radius of noticeable primary C-line and its location in nonlinear medium on the initial ellipticity degree $M_0$ and the parameter P. The points corresponding to the same values of $M_0$  are joined in lines for the sake of visibility; the ends of the lines are marked by circles. Four different values of parameter $\mu$ are considered: $\mu=0$ (a, b, c, d), $\mu=0.01$ (e, f, g, h),  $\mu=0.07$ (i, j, k, l),  $\mu=0.3$ (m, n, o, p). Diagrams (j) and (k) contain transparent dark areas, which depict the set of parameters $M_0$ and $P$ for which it's impossible to determine a single point of the primary singularity. Pictures (i) and (l) contain black dashed lines that connect the gaps caused by such uncertainty.}
\label{diagrams}
\end{figure}

It can be seen from the above results that several $C$-lines can be formed during the propagation of radiation in a ChLC. The radiation intensity at the points of polarization of singularities belonging to different $C$ lines may be comparable to the intensity at the center of the beam, or be significantly lower. Thus, the ratio of the intensity at the singularity point to the intensity at the center of the beam in the same cross section makes it possible to judge the visibility of each singularity. Our experiment showed that the brightest is the singularity formed near the input boundary of the medium (blue circles in Fig. \ref{fig:my_label}). Moreover, it is these singularities that are most stable and persist over a wide range of parameters. In what follows, we will call these singularities ``primary'', as it is done, for example, in work \cite{Grigoriev:s}. As in this work, to study the influence of the parameters of the medium and radiation on the polarization singularities, we introduce the dimensionless parameter $\gamma_{\pm}$ for both radiation components, which is equal to the ratio of the intensity at the singularity point, respectively, of the right- and left-hand polarized components to the intensity on beam axis in the same cross section:
\begin{equation}
\gamma_{\pm}=\frac{ |u_{\pm}(r_s, z_s)|^2 }{ |u_+(0, z_s)|^2+|u_-(0, z_s)|^2 },
\end{equation}
where $r_s$,$z_s$ are normalized singularity coordinates. In contrast to studies mentioned above, our work considers a chiral medium. Therefore, to study the influence of the main parameters of the medium and radiation on the nucleation of primary singularities of both polarization components, we constructed diagrams that reflect the value of the parameter $\gamma$ at the singularity points of both polarization components for the investigated range of parameters $M$ and $P$. Such diagrams were created for four different values of the $\mu$ parameter. The results are shown in the figure \ref{diagrams}. From the diagrams obtained, it becomes clear how with an increase in the value of the parameter $\mu$, which characterizes the degree of nonlocality of the nonlinear optical response, the symmetry in the distribution of the parameter $ \gamma $ is broken. Thus, as the value of the parameter $\mu$ increases, the off-axis singularities of the left-hand polarized component become more and more noticeable against the background of the intensity on the beam axis, while the singularities of the right-hand polarized component cease to be distinguishable (situation is the opposite in case of negative values of $\mu$). This phenomenon is undoubtedly associated with the presence of the chirality of the medium, which exerts a stronger focusing effect selectively on one of the polarization components of the radiation at larger values of the parameter $\mu$. Figures \ref{diagrams}(a, d), (e, h), (i, l), (m, p) show the destruction of symmetry as the value of $\mu$ grows as well. They depict the dependencies of the radius of noticeable primary C-line and its location in nonlinear medium on the initial ellipticity degree $M_0$ and the parameter P. These pictures allow us to conclude that, generally, the closest-to-axis "brightest" singularities appear at higher values of parameter $P$ (for each individual value of $M_0$). And again, they reveal the fact that with the growth of value $\mu$ the range of parameter $P$, for which primary singularities of a right-hand component are nucleated, shrinks, while the left-hand one, on the opposite, is growing.

One of the most interesting facts that we discovered, however, is the following. We found out that at some points on fig.3(j) and 3(k) it is impossible to determine which singularity we should consider as the primary one. This uncertainty is caused by the fact that at some point in dimension of parameters ($M_0$,$P$) the primary singularity divides. Two more singularities (of opposite charges) are nucleated in such points. The areas, where there do exist more than one singularity that were nucleated from the primary one, are marked by the dark areas. On the borders of these areas the nucleation of two singularities from the primary one appears (annihilation, if we consider opposite direction - going outside the dark area). Their position was defined experimentally. This hysteresis-like process is of special interest due to the next level of uncertainty that has been discovered for the first time. It is yet to find out experimentally if is possible that not only one pair of additional singularities is being nucleated in such areas, but more of them. However, theoretically this process seems very likely.

\section{Conclusions}
During the propagation of an initially uniformly polarized Gaussian beam in the bulk of a ChLC, the formation of a multifocal structure of radiation can be observed. In this case, a change in the direction of rotation of the electric field vector to the opposite leads to a significant change in light propagation dynamics, which is not observed in nematic LC. When propagating over a distance of several diffraction lengths, the ellipticity degree on the beam axis begins to tend to a certain constant value, however, regimes in which this value periodically fluctuates at distances exceeding three diffraction lengths are also observed. During the self-focusing of radiation, circular polarization singularity lines are formed in certain separated cross sections. As a rule, $C-$lines are generated in cross sections, close to the local maxima and minima of the intensity of the corresponding circularly polarized component in the center of the beam. In addition, as nonlocality of nonlinear effects is small, the primary singularities of the right- and left-polarized components are formed in an approximately symmetric space of the initial ellipticity degrees and powers of the incident radiation. However, with an increase of the absolute value of the liquid crystal chirality parameter this symmetry is broken, and, at the same values of the initial ellipticity degree, the primary singularities of one of the radiation components are formed in a smaller range of incident radiation powers, and the visibility of these singularities decreases. 
We showed that closest-to-axis "brightest" singularities appear at higher values of parameter $P$ (for each individual value of $M_0$). We also found out that as value of $\mu$ increases, the range of parameter $P$, for which primary singularities of a right-hand component are nucleated, shrinks, while the left-hand one, on the opposite, is growing.
We discovered that at some points in ($M_0$,$P$) dimension it is impossible to determine which singularity we should consider as the primary one. This uncertainty is caused by the fact that at some point in dimension of parameters ($M_0$,$P$) the primary singularity splits into two separate ones, which travel through the cross-section of the beam in different ways, but, usually, meet again and annihilate. Such points form vast areas. It seems possible that not only one pair of additional singularities is being nucleated in such areas, but more of them.

\section*{Acknowledgements}
This work was supported by the Foundation for the Advancement of
Theoretical Physics and Mathematics ``Basis''.

\bibliographystyle{ieeetr}
\bibliography{references}
\end{document}